\documentclass[a4paper]{llncs}

\usepackage{graphicx}
\usepackage{url}

\usepackage{eurosym} 
\usepackage{color}
\usepackage{euro}
\usepackage{paralist}

\graphicspath{ {./images/} }

\renewcommand{\note}[1]{$^{(#1)}$}
\usepackage[utf8]{inputenc}
\usepackage{newunicodechar}
\newcommand{\textprime}{\ensuremath{'}}
\newunicodechar{′}{\textprime}
\usepackage{setspace}

\usepackage{enumitem}
\usepackage{comment}

\begin{document}

\title{Privacy Impact Assessment: Comparing methodologies with a focus on practicality}

\author{Tamas~Bisztray\orcidID{0000-0003-2626-3434} \and
	Nils~Gruschka\orcidID{0000-0001-7360-8314}}
\institute{Department of Informatics, University of Oslo, Norway \\ 
\email{\{tamasbi$|$nilsgrus\}@ifi.uio.no}}

\maketitle

\begin{abstract}
Privacy and data protection have become more and more important in recent years since an increasing number of enterprises and startups are harvesting personal data as a part of their business model. One central requirement of the GDPR is the implementation of a data protection impact assessment for privacy critical systems. However, the law does not dictate or recommend the use of any particular framework.

In this paper we compare different data protection impact assessment frameworks. We have developed a comparison and evaluation methodology and applied this to three popular impact assessment frameworks. The result of this comparison shows the weaknesses and strengths, but also clearly indicates that none of the tested frameworks fulfill all desired properties. Thus, the development of a new or improved data protection impact assessment framework is an important open issue for future work, especially for sector specific applications. 
\footnote{
This paper is an updated version of DOI:10.1007/978-3-030-35055-0\_1 and
includes language enhancements. The changes in no case have affected the
paper’s scope, analysis, and derived conclusions. The original version of this
research is included in the proceedings of the 2019 Nordic Conference on Secure IT Systems (NORDSEC).
Editor: Tamas Bisztray.
Date: October 2021
}

\end{abstract}
\section{Introduction} 
Data is the new currency of the 21${th}$ century which we use to pay for several online services, including but no restricted to social media. There is an increasing number of businesses collecting and storing our personal information and making monetary benefits from it. The key to success for these businesses is to harness value from the collected data. To do this they need not just to store but to process what has been collected. Monetary benefits and business goals are easily pushing the protection of peoples' personal privacy down in the to-do list. The General Data Protection Regulation (GDPR) \cite{european_parliament_&_council_regulation_2016} was meant to give back the control to individuals over their private data.


Companies failing to fulfill requirements imposed by the GDPR can face serious fines laid out by Article 83, which in the worst case can be up to \EUR{20} $million$, or $4\%$ of the firm’s worldwide annual revenue from the preceding financial year. 

If a company wants to avoid such high fines they have to adjust their operations to become compliant with the GDPR. However, many worry that these regulations are imposing an unnecessary burden on tech companies. This opinion was also voiced by Alibaba's founder Jack Ma who said in an interview: ``Europe will stifle innovation with too much tech regulation''\cite{zen_soo_alibabas_2019}. Indeed for a startup such a fines would be devastating and allocating too much resources to become fully compliant could be similarly harmful. Therefore, becoming  GDPR compliant from the beginning without too much hassle is very important. But more than one year after the GDPR came into force there still exists no standard framework in the EU, and companies are either doing assessment on their own or they have to find out which DPIA framework is the one that would suit their project the best.

Article 6(1) contains six requirements one of which is necessary to fulfill in order to lawfully process PII (personally identifying information). One possibility to a obtain legal basis is to get consent from the data subject. Article 7 states that consent shall be requested in a way that is: \textit{``clearly distinguishable from the other matters, in an intelligible and easily accessible form, using clear and plain language''}

What is intelligible easy to understand plain language or what is appropriate in length can be matter of debate. Consent, once acquired is very easy to demonstrate and allows the controller\footnote{Natural or legal person, public authority, agency or other body which, alone or jointly with others, determines the purposes and means of the processing of personal data.} to specify all processing purposes to minimize the possibility of unwanted legal consequences, whereas other requirements could be more difficult to demonstrate or obtain.

It could seem like acquiring compliance can be easily achieved by listing an extensive list of processing operations in a sophisticated way and if the user clicks accept the processing is lawful. 
Google's case serves as a counterexample where the company received a \EUR{50} million fine, inflicted by French data protection authorities (CNIL) \cite{kristof_van_quathem_google_2019}. According to CNIL the way Google obtained consent violated the transparency of obligations and was lacking legal bases.
This case sets a good example but unfortunately, in practice there are still many instances where, before giving consent the user is presented with a sophisticated document in which copyright claims and other legal matters are mixed with processing purposes. 

There are several rules the data controller has to follow. Following our example with consent Article 7(3) says ``\textit{The data subject shall have the right to withdraw his or her consent at any time}.'' If the data subject withdraws consent for storing their personal information the data controller shall comply. However, in practice this usually doesn't mean immediate deletion and personal data can be retained for extended periods, which is a serious privacy risk. If the data was copied, shared with several processors, has been processed after which new PII was created, all of this should be deleted as well. This, on the data controller's side requires the ability to accurately keep track of the personal data in connection to the data subject, which is not a straightforward task.

Obtaining and maintaining compliance with the GDPR requires attention and a good overview of processing operations related to PII. Improper management of PII can lead to the violation of the GDPR. This is specially challenging for companies with complex systems designed prior the GDPR came into force. Tackling this problem requires a methodological approach, guiding organisations to identify and document activities related to PII and to assess risks related to its processing. Furthermore, it has to aid in achieving compliance with the GDPR and provide the ability to demonstrate it. 

To tackle this and other challenges, Article 35 of the GDPR introduces the concept of DPIA (data protection impact assessment). It's a method that provides guided steps to identify and analyse how the rights and freedoms of individuals might be affected by certain actions or activities related to data processing, and to assist in avoiding/correcting these issues.

This paper will analyse and compare three DPIA frameworks. The goal of this work is twofold. Firstly, we are aiming to identify shortcomings of the selected frameworks. We will do this by proposing a metric that helps to identify advantages and disadvantages of these frameworks and which could be applicable to others. The second goal is to provide help for those who are planning to conduct a DPIA, but can't decide which framework would be the best for their application as of today, and most importantly be aware of the shortcomings of the method of their choice. To help with this question we are briefly going through the selected three frameworks, examining their steps while highlighting their strengths and weaknesses.

Sections 2 gives a general introduction to DPIA and an overview of the examined DPIA frameworks. Section 3 provides an overview of related work in comparing DPIA frameworks. In Section 4 we present our metric for comparing DPIA frameworks and perform the comparison itself. Section 5 contains the summary and conclusions.

\section{Data Protection Impact Assessment}

\subsection{Legal background}
Data protection impact assessment is a requirement under the GDPR as part of the \textit{data protection by design and by default} principle (introduced in Article 25). According to Article 35:

\begin{quote}
    \itshape
    If the processing is likely to result in a high risk to the rights and freedoms of natural persons, the controller shall, prior to the processing, carry out an assessment of the impact of the envisaged processing operations.
\end{quote}

\noindent
Unfortunately, the GDPR does not specify which type of processing requires a DPIA. The reader might be confused right from the start after finding these methodologies under the name of Privacy Impact Assessment (PIA). In practice these are often used interchangeably. Originally PIA was is aiming to assess the privacy risks of a processing operations, whereas the GDPR requires DPIA to determine sufficient security measures and safeguards to mitigate risks to data subject rights rather than to mitigate privacy risks. As pointed out by Roger Clarke \cite{roger_clarke_roger_2016} in his \textit{Comprehensive Interpretation of Privacy}, data privacy is just one of the four aspects of privacy where the other three are: privacy of the person, privacy of personal behaviour and privacy of personal communications. In this paper we don't differentiate between PIA and DPIA as the \textit{"Guidelines on Data Protection Impact Assessment"} published by the Article 29 working party (WP29) recommends several PIA frameworks to be used when conducting a DPIA. 

The guideline lists nine types of processing operations likely to result in a high risk scenario (and therefore requiring a DPIA) \cite{article_29_working_party_guidelines_2017}. Similarly, there is a list of cases in which no DPIA is required. The general guideline is that if the controller is unsure where a certain process requires a DPIA, it has to be conducted. 
Furthermore, Article 35(11) requires that a new DPIA has to be carried out when there is a change in the risks related to the processing operation. This means that processes must be tracked over time in order to detect these changes. This also applies to processes which are at the moment labeled low risk since consequences and harms related to a processing operation can negatively change based on other circumstances such as time or technological advancements.



The GDPR does not reference a concrete DPIA framework, but Article 35(7) defines the basic requirements of a DPIA:

\begin{itemize}\itshape
    \item a systematic description of the envisaged processing operations and the purposes of the processing, including, where applicable, the legitimate interest pursued by the controller;
    \item an assessment of the necessity and proportionality of the processing operations in relation to the purposes;
    \item an assessment of the risks to the rights and freedoms of data subjects referred to in paragraph 1;
    \item the measures envisaged to address the risks, including safeguards, security measures and mechanisms to ensure the protection of personal data and to demonstrate compliance with this Regulation taking into account the rights and legitimate interests of data subjects and other persons concerned.
\end{itemize}

\noindent
Therefore, these points must be integrated into all DPIA frameworks aiming to assist in reaching compliance with the regulation.

WP29 recommends four generic DPIA frameworks from the EU (DE, ES, FR, UK) \cite{article_29_working_party_guidelines_2017} and one international (ISO/IEC 29134).
Additioanlly, two sector-specific frameworks \textit{``Privacy and Data Protection Impact Assessment Framework for RFID Applications''}\cite{european_commission_privacy_2011} and \textit{``Data Protection Impact Assessment Template for Smart Grid and Smart Metering systems''}\cite{european_commission_smart_2014} are recommended. 

To perform our DPIA evaluation, we selected DPIA frameworks from this list. Additionally, we will use LINDDUN as framework developed for privacy threat analysis. We used the following criteria when selecting the frameworks to compare:
\begin{itemize}
    \item The method has been recently updated.
    \item A version in English is available.
    \item The method offers a good selection of supporting material
    \item All of the following categories must be represented by one of the frameworks: policy-driven, academic, international
\end{itemize}
Based on these criterias, the following frameworks have been selected for analysis and comparison: LINDDUN, CNILs PIA framework, ISO/IEC 29134:2017. In the following sections these frameworks are presented in more detail. Throughout the analysis remarks will be made in the form of numbered notes to underline positive or negative aspects of each framework. These will be referenced in the evaluation.


\subsection{LINDDUN}

LINDDUN is a privacy threat analysis framework, developed by researchers from the DistriNet Research Group at KU Leuven, Belgium \cite{wuyts_linddun_2015}. LINDDUN, is an acronym for Linkability, Identifiability, Non-repudia\-tion, Detectability, Disclosure of information, Unawareness, Non-compli\-ance and consists of 6 main steps.

\subsubsection{1. Define Data Flow Diagram (DFD):}
As a first step data flow diagram has to be prepared to provide a high level graphical description of the whole architecture (based on SDL threat modeling). In a system model there are 4 different building blocks: entity, process, data flow, data store.

 
\textit{Note 1:} This representation makes it possible to differentiate between threat analysis for incoming and outgoing information and privacy analysis on internal data flow and processes. It is also useful to map and visualize existing architecture on a higher level.

\subsubsection{2. Mapping Privacy Threats to DFD:}
\begin{figure}
\includegraphics[scale=0.50]{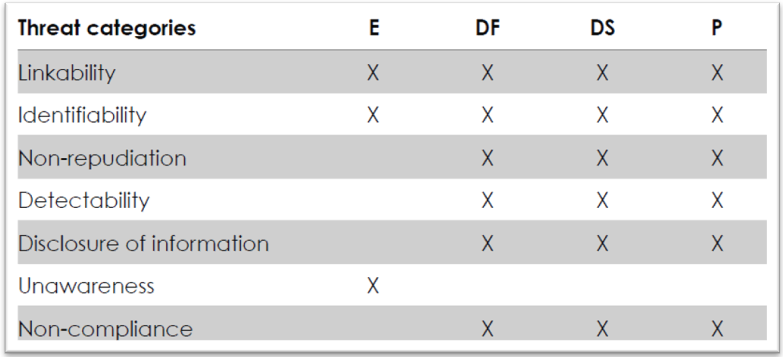}
\centering
\setlength{\belowcaptionskip}{-20pt}
\caption{Mapping privacy threats to DFD object types (E-entity, DF-data flow,
DS-data store, P-process) (Source: \cite{wuyts_linddun_2015})}
\label{fig:linddun}
\end{figure}

This step helps to identify threats connected to each object in the DFD. Figure \ref{fig:linddun} shows the LINDDUN mapping table, where each row is a LINDDUN threat category and the columns contain all DFD objects types. If a threat is relevant to a DFD object it is marked with X.

\subsubsection{3. Identify Threat Scenarios:}
Each X in the table has to be examined to determine whether the threat category is applicable to the object. Lindunn provides a set of privacy threat threes to each combination of threat category/DFD-object meaning, one for each X with the exception of Disclosure of Information, where LINDDUN points to STRIDE for further analysis. If the threat is relevant, threat scenarios has to be documented from the threat actors point of view. Otherwise the assumptions on why something is not relevant has to be documented.

\subsubsection{4. Prioritise Threats:}
There can be an overwhelming number of threats and due to budget and time limitations threats must be prioritised and reviewed in the order or relevance. Risk assessment is not part of LINDDUN and it references a number of external frameworks to perform this step: OWASP's Risk Rating Methodology, Microsoft's DREAD, NIST's Special Publication 800-30, or SEI's OCTAVE.\newline
\noindent\textit{Note 2:} 
Pointing to various external resources can make it difficult for the controller to review and pick the one most suitable for their case. Including risk assessment tailored to LINDDUN would improve practicality.

\subsubsection{5. Elicit mitigation strategies:}


Every threat tree is automatically linked to a mitigation strategy (which later is linked to solution steps). This can create a lot of unnecessary steps and could hinder the analyst from considering strategies no presented. As certain mitigation strategies can be applicable to multiple points, providing an extensive list of strategies can be also useful without direct connection to threat trees.

In a case study where LINDDUN was used for DPIA with a Identity Wallet Platform \cite{veseli_engineering_2019} the authors considered this impractical and used the ISO/IEC 27005 \textit{Information Security Risk Management}, which identifies four mitigation strategies: risk reduction, retention, avoidance, and transfer.

\subsubsection{6. Select corresponding threats:}
For every mitigation strategy a list of related papers is provided on appropriate privacy enhancing technologies. This is organised as a table and can be found in the supporting materials. In \cite{veseli_engineering_2019} the authors noted that they faced “lack of expertise, low technology readiness level, and other uncertainties regarding the integration of Privacy-Enhancing Technologies (PETs)”. They also identified further PETs not present in the table. They also described the need of finding balance between project goals and privacy goals as they had trouble addressing all privacy threats.

\subsection{CNIL} 

The CNIL PIA framework was created by the French data protection authority \cite{commission_nationale_de_linformatique_et_des_libertes_privacy_2019}. It mainly focuses on GDPR compliance and the guidelines of WP29 \cite{article_29_working_party_guidelines_2017}. It has a well rounded list of supporting materials: Methodology, Knowledge bases, Templates, Application to IOT devices, examples of data processing operations likely to result in a high risk \cite{commission_nationale_de_linformatique_et_des_libertes_analyse_2018}, and a software tool that helps to go through the steps \cite{commission_nationale_de_linformatique_et_des_libertes_open_2019}.
It helps to demonstrate compliance and provides guiding steps to achieve it. Here, compliance is defined as a combination of the following two pillars: Fundamental Rights and Principles (non-negotiable), Management of data subjects/privacy risks (technical controls).
It consists of 4 steps: describe context of processing, guarantee compliance with fundamental principles, assess privacy risks and treat them, document validation. 

\subsubsection{1. Study of context:}
The main steps here are:
\begin{itemize}[noitemsep,topsep=3pt]
    \item outline processing
    \item identify data controller and any processor
    \item check applicable references: approved codes of conduct (Article 40), certifications regarding data protection (Article 42)
    \item define personal data concerned / recipients, storage duration
    \item describe processes and personal data supporting assets
\end{itemize}

\textit{Note 3:} As a first step it doesn't give a good picture of the whole architecture, data movement is not visible and doesn't allow to later differentiate data protection and privacy analysis. The online tool doesn't support grouping or any structuring of this information. On the upside it references the GDPR and the relatable sections to help with compliance and clearly defines what is expected at each entry and what the data controller should pay attention to.

\textit{Note 4:} Article 35(1) requires “a systematic description of the envisaged processing operations”. This could be a matter of debate but in contrast to LINDDUN or ISO, CNIL doesn't fully comply with this point. It does list and describe the processing operations but the process itself doesn't aid the controller in a systematic way.

\textit{Note 5:} It's important to remark that CNIL in its collection of supporting materials among which is a document called ``Templates", does provide tables to collect and categorise information. It is not as good as a visual representation but good for record-keeping. However, this paper focuses on the process and the practicality of the DPIA framework. If during the steps of the DPIA it is not explicitly mentioned or referenced if something additional is needed for that very step, or the concept is not present in the main document which describes the process, it will not be counted as fully part of the process. A DPIA should be intuitive with sufficient guidance provided. Due to the plethora of templates available it is not easy to get a hold of what is needed for a certain step. The main document itself never references to any of the templates and their lack of integration into the process is a serious drawback.

\subsubsection{2. Study of fundamental principles:}
The aim is to ensure compliance with privacy protection principles. It consists of two steps: \textit{``Assessment of the controls guaranteeing the proportionality and necessity of the processing to enable the persons concerned to exercise their rights"} and \textit{``Assessment of controls protecting data subjects' rights"}.


\textit{Note 6:} This step is a direct translation of the second bare minimum principal from Article 35 (7b) and it mainly focuses on a legal compliance rather than privacy and data protection principles.


\subsubsection{3. Study of Risk:}

\textit{Note 7}: So far the software tool and the written material were in sync, the tool used the same points described in the text but from this point on it forks into two different processes with different questions.

The document first gives a general introduction to risk assessment with a brief overview on how to calculate risk level. The supporting material \textit{``Knowledge Bases''}  provides a very detailed tutorial on how to determine severity and likelihood with a lot of real life threat scenarios. Study of risks contains two sections. 

The first one is 
\textit{``Assessment of existing or planned controls''} on controls of data being processed: encryption, anonymization etc., general security controls, and organisational controls. \textit{Note 8:} The order of the steps so far are incorrect in the documentation. Potential threats were not yet identified neither controls mitigating those threats. Encryption is used if for example confidentiality needs to be protected, but there are many forms of encryption and first a threat needs to be recognised to use the proper countermeasures.

The second section is on \textit{``Risk assessment''.} It requires the following steps for each impending event:
\begin{itemize}[noitemsep,topsep=3pt]
    \item determine potential impact on the data subject privacy
    \item estimate severity of impact
    \item identify threats to personal data supporting assets, that leads to this feared event (threat scenario) and the risk sources (threat actors)
    \item estimate likelihood
    \item determine whether the risks identified in this way can be considered acceptable in view of the existing or planned controls.
\end{itemize}

\textit{Note 9:} These steps are out of order. Likelihood and severity has to be calculated before impact. \textit{Note 10:} Considering if a risk can be acceptable doesn't qualify as prioritising threats.

The software tool follows a different path: Planned or existing measures, illegitimate access to data, unwanted modification of data, data disappearance, and risks overview. It eerily resembles the CIA triad (confidentiality, integrity, availability) with \textit{``Planned or existing measures''} and \textit{``Risk overview''} added. It also doesn't categorise the risk properly, neither differentiates between non-negotiable ones and technical controls.

\subsubsection{4. Validation of the DPIA:}
In a timely fashion this section correspond to Article 35 (7d): ``the measures envisaged to address the risks, including safeguards, security measures and mechanisms to ensure the protection of personal data and to demonstrate compliance...''. It consists of 3 steps: prepare material, formal validation, repeat when necessary (no further comment on how often).

\subsection{ISO/IEC 29134:2017}
The ISO/IEC standard number 29134:2017 also has a good selection of supporting material, mainly other ISO standards like a risk-based management system: ISO/IEC 27001, or overview and vocabulary: ISO/IEC 29100:2011 etc. The document itself starts with a long discussion on principles and guidelines related to conducting the DPIA such as: preparing grounds, benefits of DPIA, objectives of reporting, accountability to conduct, scale, determine if DPIA is necessary, preparations, set up a team, prepare a plan, describe what is being assessed, and stakeholder engagement.

The actual steps of the DPIA only starts at Section 6.4 and it consists of 5 main steps:
\begin{enumerate}[topsep=3pt]
    \item Identify information flows of personally identifying information (PII)
    \item Analyse the implications of the use cases
    \item Determine the relevant privacy safeguarding requirements
    \item Assess privacy risks
    \item Prepare for treating privacy risks
\end{enumerate}

The general structure for these steps consists of the following list the conductor has to fill out:
\textit{Objective},
\textit{Input},
\textit{Expected output},
 \textit{Actions},
 \textit{Implementation guidelines}.


The guiding document provides a detailed description of what is expected at each point of this list that is tailored to the main steps. After the list it provides a detailed guide with comments and recommendations to further assist the conductor of the DPIA by for example, listing organisational and non-compliance threats and other tips related to that part of the assessment.
\textit{Note 14:} This approach makes the whole process a bit monotone almost like filling out a questionnaire but at least it draws attention to important questions. 

The next sections are the DPIA follow up and the DPIA report. These are more detailed than the DPIA itself. For example the Risk assessment section contains: Threats and their likelihood, Consequences and their level of impact, Risk evaluation, Compliance analysis. But it always references back to previous points so it doesn't mean it was missing from the process. Plus it is easy to put together and provides a very detailed report.

\section{Related Work}
Although there is an overabundance in available DPIA frameworks there hasn't been a lot of work on evaluating and comparing them. Further, from the existing literature only a small proportion was published after the GDPR came into force. There are two types of papers in this topic. Evaluating DPIA frameworks and comparing/measuring the effectiveness of DPIA reports. These are two different fields but in the pursuit of evaluation a lot can be learned from the study of reports as well. 


As mentioned before, the GDPR unfortunately does not provide or recommend an actual framework to follow. This is a shortcoming recognised by Wright et al. already in 2013, in connection to Directive 95/46/EC, urging that the European commission and EU member states should draw from the experience of other countries and develop their own DPIA policy, methodology and framework \cite{wright_comparative_2013}. They also pointed out that a DPIA should be more than a compliance check, as it should be a process. It has to be reviewed and updated throughout the whole life cycle of the project as also stated in Article 35(11). They compared DPIA frameworks from six countries drawing inspiration from the PIAF project co-founded by the European Commission which reviewed DPIA methods from other countries \cite{piaf_privacy_2011} .

The PIAF deliverable no. 1 compared the effectiveness of DPIA guides based on a checklist of 18 questions \cite[table 10.1]{david_wright_privacy_2011} and comparing DPIA reports using checklist of 10 questions (no. 2) \cite[table 10.2]{david_wright_privacy_2011}. The third deliverable  outlines what a DPIA process should contain \cite{de_hert_recommendations_2012}. These are: Project description, Stakeholder consultation, Risk management, Legal Compliance check, Report, Implementation, Review.


The RFID framework \cite{european_commission_privacy_2011} even though it is sector-specific, recognises eight important steps a DPIA should address. These are: characterization of the application, initial analysis, definition of privacy targets, evaluation of degree of protection demand for each privacy target, identification of threats, identification and recommendation of controls, consolidated view of controls, assessment and documentation of residual risks. The frameworks our analysis will focus on are those designed for general use-cases.

So far, the most commonly used technique for comparing DPIA frameworks was to go through a checklist to see if it fulfills certain requirements. However, this is most effective if the evaluation criteria is quantitative in nature and only checks the existence or non existence of a certain aspect. For example in the comparison of Wright at al. one checkpoint is: ``Provides a suggested structure for the PIA report''. This is a binary question.  Whereas other qualitative matters shouldn't be written off with a check-mark. For example points like \textit{DPIA is a process} or \textit{DPIA is more than a compliance check}. It doesn't matter if a framework claims to accomplish these points, the real question is how well they fulfill these requirements. It is also pointless to include such questions in a DPIA report analysis as it is not the job of the project owner to perfect the DPIA framework to have a better workflow, rather it is the task of those developing DPIA frameworks and tools to fulfill these requirements. A report is a summary on the results and the findings from conducting the DPIA. These problems are commonly present in previous works by either treating important question as a check-mark and not uncovering shortcomings of the DPIA process, or including questions in the DPIA framework comparison related to the quality of the output of the framework which is the report analysis and not the analysis of process itself. 

An improvement to the check-mark approach was PEGS (PIA Evaluation and Grading System) proposed by Whadwa et al. \cite{wadhwa_evaluating_2013}. Even though this method was developed to evaluate the actual DPIA process post facto (the DPIA report) not the DPIA itself, the authors note that it can be helpful also in guiding a DPIA process. Their evaluation criteria is first presented as a checklist where they provide an extra column, where in case a requirements was not fulfilled \textit{scope of improvement} can be specified. Then a weighting is applied (1, 2 or 3) to each criteria in line with their relative importance, where 1 is the least important and 3 is the most important. The choice of assigning the weights was highly based on the PIAF project. Criteria with weight 1 includes: clarification of early initiation, identification of who conducted the DPIA and publication; weight 2: project description, purpose and relevant contextual information, information flow mapping, legislative compliance checks and identification of stakeholder consultation; weight 3: identification of privacy risks and impacts, identification of solutions/options for risk avoidance and mitigation, and recommendations handling after the PIA.

In a more recent analysis Vemou et al. \cite{vemou_evaluation_2018} reviewed 9 different DPIA Frameworks regardless of country of origin with the only criteria that it should not be sector specific, by using 17 questions derived from existing literature as check-mark points to draw attention the to lack of completeness of these frameworks.

\section{Comparison of DPIA frameworks}
\subsection{Comparison Metric}
To successfully evaluate a DPIA framework we should differentiate between three categories. Questions that relate to steps someone should consider prior starting the DPIA are preliminary questions, they are important to consider but they are not part of the core process. Similarly, there are a series of questions only relevant after a DPIA is completed for example: “is the publication of the DPIA report provisioned”. These are in the territory of DPIA report analysis (which is  the scope of our future research). The questions that we focus on are the ones directly related to the core points of a DPIA which is based on Article 35 of the GDPR. Questions should also point out the shortcomings identified during the review of the examined frameworks remarked as \textit{Notes}. 

Some questions such as “is it a process” or “is structured guidance to assist in risk assessment provided?” cannot be answered with a single check-mark as they are rather qualitative questions. However, it would be very difficult to build an evaluation framework with open questions. In previous works questions related to the DPIA process were simply listed. Here the questions will be structured based on which part of the DPIA process that question should belong to. It's important to note that Article 29 working party's document contains the only official documentation on how the main steps of a DPIA process should look like. Therefore, we consider that as a starting point. Their \textit{``Criteria for acceptable DPIA''} checklist however, cannot be taken as a blueprint when it comes to the internal part of these main steps. The points of Article 35 (7) of the GDPR was also only meant to be a list. Unfortunately, the steps of CNIL for example, structures the internal parts of these main steps as it was presented in these documents.

In total we collected 28 questions. These are categorised based on which part of the DPIA they belong to. The evaluation happens the following way. Each questions will receive grades in two categories: \textit{score} (S) and \textit{process} (P). The \textit{score} (S) meant to determine if the question is covered by the examined DPIA framework. For the \textit{score}, a question can get 0 points if it is not in covered by the framework, 1 point if it is partially included, and 2 if it is extensively addressed. The \textit{process} (P) meant to evaluate if a question is well integrated in the DPIA process. For example it is discussed in the right part of the DPIA, and not just mentioned by the framework without merging it logically from the perspective of the whole process. It must be properly discussed at the right time, with adequate references and supporting materials if necessary. The \textit{process} (P) can receive a grade of $+1$ if the step is integrated into the process, or $-1$ if it is not integrated. If the \textit{score} is zero (question not in covered by the framework), the \textit{process} is automatically zero too. This approach can penalize if a framework while mentioning certain criterions or questions, doesn't integrate the steps into a process and it's closer to a compliance check. For a simple check-mark evaluation this is not possible.

In this analysis we are not going to determine whether the examined framework supports full compliance, but several of these questions are going to assess this criterion. Checking all the applicable points of the GDPR is not the scope of this paper, as it is more important during the report evaluation.

\subsection{Evaluation Questions}
This section contains the grading questions split up between six tables for the six steps we identified as crucial parts of the DPIA process. The cells apart from the received grades in some cases also contain a reference to the \textit{``Notes"}. For example \textit{Note 1} is denoted as \note{1}. Grades for (S) and (P) are also separately shown \textit{(Breakdown)} before summarised in the \textit{Total} score where the maximum achievable grade is also shown. Most questions are aiming to evaluate the content of the frameworks, while others are specifically trying to uncover if the order of steps and questions are designed properly.

\hfill
\break
\begin{tabular}{ |p{8.5cm}|p{5mm}|p{5mm}|p{5mm}|p{5mm}|p{5mm}|p{5mm}| }
\hline
Step 1: Description of envisaged processing & \multicolumn{2}{|c|}{ISO} & \multicolumn{2}{|c|}{CNIL} & \multicolumn{2}{|c|}{LIN.}\\
& S & P & S & P & S & P \\ 
\hline\hline
Structured description and mapping of information flows, contextual information and envisaged processing (structured: either graph or table) & 2 & +1 &2 & -1 \note{3} \note{4} \note{5} & 2 & +1 \note{1}\\ 
\hline
Establish easy to follow connections between system elements (data, process, supporting assets etc.) &  2 & +1 & 1 & -1 \note{5}& 2 & +1 \note{1}\\
\hline
Allow the differentiation of internal and external data movement & 2 & +1 & 0 & 0 &  2 & +1 \note{1} \\
\hline
Stakeholder identification  & 2 & +1 & 2 & +1 &  1 & +1\\
\hline
\hline
Breakdown & 8 & +4 & 5 & $-1$ & 7 & +4 \\
\hline
Total score (out of 12) & \multicolumn{2}{|l|}{12} & \multicolumn{2}{|l|}{4} & \multicolumn{2}{|l|}{11}\\
\hline
\end{tabular}
\hfill 
\break
Both LINDDUN and ISO are using visual representation of the information flows and they are described during the process. LINDDUN is more intuitive but the instructions provided in ISO are more detailed. Unfortunately, the CNIL framework really falls behind at this point. This step is the foundation stone for the whole DPIA and missing points here can jeopardize the effectiveness of the whole process.
\hfill 
\break
\hfill
\break
\begin{tabular}{ |p{8.5cm}|p{5mm}|p{5mm}|p{5mm}|p{5mm}|p{5mm}|p{5mm}| }
 \hline
Step 2:  Assess necessity and proportionality of processing & \multicolumn{2}{|c|}{ISO} & \multicolumn{2}{|c|}{CNIL} & \multicolumn{2}{|c|}{LIN.}\\
 & S & P & S & P & S & P \\ 
 \hline\hline
How information is to be collected, used, stored, secured and distributed and to whom and how long the data is to be retained & 2 & +1 & 2 & +1 & 0 & 0 \\ 
 \hline
Compliance with Article 29 Working party's corresponding list (Annex 2/necessity and proportionality) &  2 & +1 & 2 & +1 & 0 & 0\\
 \hline
 Analyse all previously identified system elements in a structured manner & 2 & -1 & 0 \note{6} & 0 \note{6}& 2 & +1 \\
\hline
  \hline
  Breakdown & 6 & +1 & 4 & +2 & 2 & +1 \\
   \hline
 Total score (out of 9) & \multicolumn{2}{|l|}{7} & \multicolumn{2}{|l|}{6} & \multicolumn{2}{|l|}{3}\\
   \hline
\end{tabular}
\hfill
\break

LINDDUN is missing legal assessment and mainly focuses on threat analysis. CNIL does a good job from a compliance perspective but it doesn't connect the dots. We deviate from the steps of WP29 where the following would be: "measures envisaged". To determine organisational and technical is important to know the context, and the threaths/risks. Here we prefer the approach of LINDDUN where an early threat analysis is initiated.
\hfill
\break
\hfill
\break
\begin{tabular}{|p{8.5cm}|p{5mm}|p{5mm}|p{5mm}|p{5mm}|p{5mm}|p{5mm}|}
\hline
Step 3. Identify threats/risks & \multicolumn{2}{|c|}{ISO} & \multicolumn{2}{|c|}{CNIL} & \multicolumn{2}{|c|}{LIN.}\\
& S & P & S & P & S & P \\ 
\hline\hline
Organisational and technical that are endangering the rights of data subjects & 2 & +1 & 2 & +1 & 0 & 0 \\ 
\hline
Origin of risks are specified (threat actor-attack surface)  & 2 & $-1$ & 2 & $-1$ &  2 & +1 \\
\hline
Threats should be directly linked to elements from the first step &  1 & $-1$ & 1 & $-1$ & 2 & +1\\
\hline
Identification of threats coming from GDPR non-compliance is integrated into the process & 2 & +1 & 2 & +1 & 0 & 0\\
\hline
Threats are identified before Risk Assessment  & 2 & +1 & 0 \note{8} & 0 &  2 & +1\\
\hline
Is there a differentiation between threat analysis and privacy analysis  & 2 & +1 & 2 & +1 &  1 & $-1$\\
\hline
Addresses all types of privacy risks (informational, bodily, territorial, locational, communications)  & 1 & $+1$ & 1 & $-1$ &  1 & $+1$\\
\hline
\hline
Breakdown & 12 & +3 & 10 & 0 & 8 & +3 \\
\hline
Total score (out of 21) & \multicolumn{2}{|l|}{15} & \multicolumn{2}{|l|}{10} & \multicolumn{2}{|l|}{11}\\
\hline
 
\end{tabular}
\hfill
\break
\hfill 
\break

The drawback of LINDDUN is again the fact that legal compliance is not integrated in the process, which the authors clearly state in the beginning, but as a result lot of aspects are missing (some are unintentionally tackled). LINDDUN only considers a limited number of threats. CNIL is very strong content-wise but there is no logical structure in the order of its steps.
\hfill
\break
\hfill
\break
\begin{tabular}{ |p{8.5cm}|p{5mm}|p{5mm}|p{5mm}|p{5mm}|p{5mm}|p{5mm}| }
 \hline
 Step 4: Risk Assessment & \multicolumn{2}{|c|}{ISO} & \multicolumn{2}{|c|}{CNIL} & \multicolumn{2}{|c|}{LIN.}\\
 & S & P & S & P & S & P \\ 
 \hline\hline
 Structured guidance to assist in risk assessment is provided & 2 & +1 & 2 & +1 & 0 & 0 \\ 
 \hline
Fundamental Rights and Principles (non-negotiable) and Management of data subjects/privacy risks (technical controls) are differentiated &  1 & +1 & 2 & +1 &  0& 0\\
  \hline
    Risk calculation is included with sufficient supporting material& 2 & +1 & 2 & $-1$ \note{9} &  0 & 0 \\
 \hline
 Risks are prioritised & 2 & +1 & 1 \note{10} & +1 &  0 & 0\\

 \hline
   Lower risks that are not immediately addressed are well documented & 2 & +1 & 1 & +1 &  2 & +1\\
 \hline
   Risk reduction, retention, avoidance, and transfer are all listed as mitigation strategies and sufficiently discussed in supporting material & 2 & +1 & 1 & +1 & 0 & 0 \\ 
    \hline
Owner of residual risks specified & 2 & +1 & 2 & +1 &  0 & 0\\
  \hline
  \hline
  Breakdown & 13 & +7 & 11 & +5 & 2 & +1 \\
   \hline
 Total score (out of 21) & \multicolumn{2}{|l|}{20} & \multicolumn{2}{|l|}{16} & \multicolumn{2}{|l|}{3}\\
   \hline
\end{tabular}
\hfill
\break

LINDDUN doens't include a risk assessment methodology, only recommends some. It receives points as in the previous step threats were documented. ISO also points to other ISO/EIC standards and guides, but it does include a structured guide on its own and the recommendations are well referenced and compatible. In contrast, LINDDUN leaves the privacy analyst alone to figure out which method would suit them the best. 
\hfill
\break
\hfill
\break
\begin{tabular}{ |p{8.5cm}|p{5mm}|p{5mm}|p{5mm}|p{5mm}|p{5mm}|p{5mm}| }
\hline
Step 5: Measures envisaged  & \multicolumn{2}{|c|}{ISO} & \multicolumn{2}{|c|}{CNIL} & \multicolumn{2}{|c|}{LIN.}\\
& S & P & S & P & S & P \\ 
\hline\hline
Technical controls and PETs are only discusses after threats and related risks have been evaluated
& 2 & +1 & 1 & $-1$ & 2 & +1 \\ 
\hline
An extensive list of organisational measures are provided &  1 & +1 & 2 & $-1$ & 0 & 0\\
\hline
An extensive and updated list of technical measures (PETs) are available & 1 & -1 & 2 & $-1$ & 2 & +1 \\
\hline
Literature/supporting material for suggested PETs are included & 0 & 0 & 2 & $-1$ & 2 & +1 \\ 
\hline
\hline
Breakdown & 4 & +1 & 7 & $-4$ & 6 & +3\\
\hline
Total score (out of 12) & \multicolumn{2}{|l|}{5} & \multicolumn{2}{|l|}{3} & \multicolumn{2}{|l|}{9}\\
\hline
\end{tabular}
\hfill
\break
\hfill
\break

CNIL again is very vague in the main document but, the fact that in it's ``knowledge bases" supporting material provides a wide selection of technical measures it get a high \textit{score}, but is heavily penalized as these are not adequately references in the process. The list provided by LINDDUN can not be considered complete (neither CNILs or ISO), but it's a step in the right direction.
 
\hfill \break
\begin{tabular}{ |p{8.5cm}|p{5mm}|p{5mm}|p{5mm}|p{5mm}|p{5mm}|p{5mm}| }
 \hline
 Step 6: Documentation/Validation  & \multicolumn{2}{|c|}{ISO} & \multicolumn{2}{|c|}{CNIL} & \multicolumn{2}{|c|}{LIN.}\\
 & S & P & S & P & S & P \\ 
\hline\hline
Outline of the report was generated during the process
 & 2 & +1 & 1 & $-1$ & 1 & $-1$ \\ 
 \hline
 Result is evaluated &  2 & +1 & 2 & +1 & 0 & 0\\
 \hline
Action plan for continuation & 2 & +1 & 2 & +1 &  1 & $-1$ \\
\hline
 \hline
  Breakdown & 6 & +3 & 5 & +1 & 2 & $-2$ \\
   \hline
 Total score (out of 9) & \multicolumn{2}{|l|}{9} & \multicolumn{2}{|l|}{6} & \multicolumn{2}{|l|}{0}\\
    \hline
\end{tabular}
 \hfill \break
 
Here ISO outruns the other frameworks in terms of DIPA report preparation. The steps are already outlined in the main document and every step is referenced back to the process. 
 
 \hfill \break
 \begin{tabular}{ |p{8.3cm}|p{5mm}|p{7mm}|p{5mm}|p{5mm}|p{5mm}|p{5mm}| }

 \hline
 Evaluation  & \multicolumn{2}{|c|}{ISO} & \multicolumn{2}{|c|}{CNIL} & \multicolumn{2}{|c|}{LIN.}\\
 & S & P & S & P & S & P \\ 
 \hline\hline
 Final Breakdown & 49 & $19$ & 42 & $3$ & 27 & $10$ \\
   \hline
\textbf{Final Score (out of 84)} & \multicolumn{2}{|c|}{68} & \multicolumn{2}{|c|}{45} & \multicolumn{2}{|c|}{37}\\
    \hline
\end{tabular}

\hspace{0.5cm}

The overall result shows that, CNIL lost a lot of points for coming off as a compliance check and not trying to be a better process. Due to the lack of references and integration of the supporting material into the main process it lost a lot of points. The ISO framework proved to be the best but it could also use improvements. Although, the order of its steps and the content are good, it feels like a questionnaire with repetitive steps. LINDDUN needs to develop a step for risk assessment and documentation. References to supporting materials and GDPR compliance must be incorporated throughout the process. 

\section{Summary and Outlook}
In this paper we performed a comparison of three popular data protection impact assessment frameworks.
By approaching the evaluation and grading from the perspective of an intuitive process rather then a compliance check, it became obvious that in many cases very important aspects of these methodologies are not properly developed. The ISO standard proved to provide the best framework both content-wise and as a process, although there are still many shortcomings waiting to be improved. The latter is even more true in case of CNIL and LINDDUN. These are among the state of the art PIA frameworks with the purpose of: helping companies implementing the Privacy by Design paradigm, support developing GDPR compliance (and to avoid fines), but mostly to assist in the protection of the rights and freedom of natural persons. CNIL has a very good selection of supporting material and in terms of achieving GDPR compliance, it is the best framework examined. However, as a genuine process it doesn't perform well. LINDDUN has a very strong start but it completely misses Risk Assessment and it's 5th step (Eliciting mitigation strategies) is cumbersome, while steps related to documentation/validation is also less developed.

Following Wright et al. \cite{wright_comparative_2013} we also highlight the importance of one or more officially approved EU-specific DPIA frameworks with sufficient and regularly updated supporting material.
In future work we will apply these frameworks to various projects to address examine how they support GDPR compliance, and analyse the DPIA reports with the intention of proposing improvements to these frameworks.

\bibliography{main}
\bibliographystyle{splncs03}

\end{document}